\documentclass[final,3p,twocolumn,sort&compress]{elsarticle}

\pdfoutput=1

\usepackage{amssymb}
\usepackage{amsmath}
\usepackage{amsfonts}
\usepackage{graphicx}
\usepackage{hyperref}
\usepackage{color}
\usepackage{xspace}
\usepackage{ulem}
\usepackage{mathtools}
\usepackage{hhline}
\usepackage{soul}

\journal{Physics Letters B}

\begin{document}

\begin{frontmatter}

\title{Towards a reliable effective field theory of inflation}

\author[aa]{Mar Bastero-Gil}
\ead{mbg@ugr.es}
\address[aa]{Departamento de F\'{\i}sica Te\'orica y del Cosmos,
  Universidad de Granada, Granada-18071, Spain}
  
\author[bb]{Arjun Berera}
\ead{ab@ph.ed.ac.uk}
\address[bb]{School of Physics and Astronomy, University of
  Edinburgh, Edinburgh, EH9 3FD, United Kingdom}

\author[cc]{Rudnei O. Ramos\corref{11}}
\ead{rudnei@uerj.br}
\address[cc]{Departamento de F\'{\i}sica Te\'orica, Universidade do Estado do
  Rio de Janeiro, 20550-013 Rio de Janeiro, RJ, Brazil}
\cortext[11]{Corresponding author}

\author[dd]{Jo\~{a}o G.~Rosa}
\ead{jgrosa@uc.pt}
\address[dd]{Departamento de F\'{\i}sica da 
Universidade de Coimbra and CFisUC, Rua Larga, 3004-516 Coimbra, Portugal}

\begin{abstract}

We present the first quantum field
theory model of inflation that is renormalizable in the matter
sector, with a super-Hubble inflaton mass and sub-Planckian field
excursions, which is thus technically natural and consistent with a
high-energy completion within a theory of quantum gravity. This is
done in the framework of warm inflation, where we show, for the first
time, that strong dissipation can fully sustain a slow-roll trajectory
with slow-roll parameters larger than unity in a way that is both
theoretically and observationally consistent. The inflaton field
corresponds to the relative phase between two complex scalar fields
that collectively break a $U(1)$ gauge symmetry, and dissipates its
energy into scalar degrees of freedom in the warm cosmic heat bath. A
discrete interchange symmetry protects the inflaton mass from large
thermal corrections. We further show that the dissipation coefficient
decreases with temperature in certain parametric regimes, which
prevents a large growth of thermal inflaton fluctuations. We find, in
particular, a very good agreement with the Planck legacy data for a
simple quadratic inflaton potential, predicting a low tensor-to-scalar
ratio $r\lesssim 10^{-5}$. 

\end{abstract}



\end{frontmatter}


Building a consistent quantum field theory (QFT) model for inflation
\cite{inflation} remains one of the great challenges of modern
cosmology. On the one hand, the lack of empirical evidence for beyond
the Standard Model particle physics motivates using a ``bottom-up"
approach, with as few novel ingredients as possible. On the other
hand, the slow-roll inflationary dynamics is extremely sensitive to
unknown physics at or even below the Planck scale, at least within the
more conventional cold inflation paradigm. Given also that in cold
inflation the simplest renormalizable scalar potentials $\phi^2$ and
$\phi^4$ have been ruled out by observation~\cite{Akrami:2018odb}, in
the last few years inflation model building has become almost entirely
``top-down", based on some putative high-energy theory such as string
theory. But this means that most models involve at least one and
usually more major new features that have not been empirically
confirmed, like supersymmetry/supergravity, special geometries,
modified gravitational couplings, non-renormalizable operators,
etc. We argue that this need not be the case, and that a reliable
inflationary model may remain close to what has been empirically
confirmed, while making predictions that are insensitive to the
high-energy completion of the model. 

One of the main problems in developing an inflation model is incorporating the light scalar inflaton field. Light scalars are extremely unnatural in any effective quantum field theory, since quantum corrections to their mass are quadratically divergent --- a symptom of their sensitivity to new physics above the cut-off energy scale below which the theory can describe physical phenomena. The Higgs boson is the paramount example of this technical naturalness problem, better known as electroweak hierarchy problem, as its mass should naturally lie close to the Planck scale if no new particle states exist below the latter (see, e.g., Refs.~\cite{Wells:2016luz, Craig2017}). According to ’t Hooft~\cite{tHooft1980}, this is related to no new symmetries emerging for vanishing scalar masses, as opposed to, e.g., fermions, for which a chiral symmetry is gained in this limit.

An effective field theory of inflation based on general relativity or any classical gravity theory necessarily fails above the Planck scale, making the scalar potential sensitive to Planck-suppressed non-renormalizable operators that generically drive the inflaton mass towards values above the Hubble scale $H$. This is the well-known "eta-problem" in e.g.~F-term supergravity and string theory (see, e.g., Ref.~\cite{Baumann:2014nda}), which are but examples of a more general inflationary naturalness problem, given that the cut-off scale for the effective field theory of inflation must necessarily lie above the Hubble scale, thus driving the inflaton mass to super-Hubble values, in tension with the slow-roll conditions. 

Many have tried to overcome this issue by employing symmetries that
could enforce cancellations between different quantum corrections to
the inflaton's mass. The best-known example is supersymmetry, where
bosonic and fermionic quantum corrections cancel out, but
supersymmetry is broken by the finite energy density during inflation,
leaving $\mathcal{O}(H)$ corrections to the inflaton mass except,
e.g., for some fine-tuned K\"ahler potentials. Global symmetries
should be broken in quantum gravity, such that shift symmetries also
cannot guarantee the flatness of the potential. This is inherently
assumed in models with axion-like fields~\cite{Freese:1990rb}, as well
as models with plateaux in the scalar potential, like Higgs
inflation~\cite{Bezrukov:2007ep} and attractor
models~\cite{Stewart:1994ts,Kallosh:2013yoa}, hence necessarily
involving some degree of fine-tuning. {}Fine-tuning is, of course,
undesirable in a theory that is supposed to dynamically generate the
otherwise extremely fine-tuned initial conditions of standard
cosmology.

There has also been an increased interest in the ``swampland
conjectures", which pose stringent conditions on effective QFTs
admitting a consistent quantum gravity completion, motivated by
explicit string theory
constructions~\cite{Ooguri:2006in,Obied:2018sgi,Ooguri:2018wrx}. In
particular, these require that  $\Delta\phi\lesssim M_{\rm P}$,
$|V'|/V\gtrsim \mathcal{O}(1)$, and more recently the transplankian
censorship conjecture (TCC) on the lifetime of the de Sitter state $t
< 1/H_{inf} \ln(M_P/H_{inf})$  \cite{Bedroya:2019snp,Bedroya:2019tba,Berera:2020dvn}.
While some inflationary scenarios can satisfy the first criteria, the
last two are in tension with inflation slow-roll and expansion
conditions.

In addition, inflationary predictions for CMB observables rely on the
consistency of QFT in curved space-time, which is also not free from
ambiguities, and for which we have no empirical guidance (see, e.g.,
Ref.~\cite{ArmendarizPicon:2003gd, Agullo:2008ka, Durrer:2009ii,
  Marozzi:2011da, Beneke:2012kn, Bastero-Gil:2013nja}). In particular,
the form of the primordial spectrum of density perturbations on
super-horizon scales depends on the particular choice of the
Bunch-Davies vacuum, which is not unique and is, moreover, sensitive
to transplanckian physics. 

It had been understood early on in the development of warm inflation
that, in the strong dissipative regime, all these crucial issues can
be overcome~\cite{Berera:1999ws,Berera:2004vm}  (see
also~\cite{Das:2018rpg, Motaharfar:2018zyb,Kamali:2019xnt,Berera:2019zdd,Brandenberger:2020oav,Berera:2020dvn}),  but
realizing this in a QFT model has proven very challenging. 
This
letter develops the first such model, which within the
matter sector, thus apart from the coupling to gravity, has only
renormalizable matter field interactions
and it is both theoretically and observationally consistent.

Dissipation is an inherent process to any system interacting with its
environment.  This has long been embodied in the warm inflation
paradigm~\cite{Berera:1995ie,Berera:1996nv,Berera:1996fm}, where it is
shown that interactions between the inflaton and particles in the
cosmic bath necessarily lead to dissipative effects and associated
particle production. Part of the inflaton's energy is thus
continuously transferred to the radiation bath, acting as a heat
source that keeps it warm despite the supercooling effect of
accelerated expansion.

Dissipative effects are encoded in the effective equation for the
inflaton field, 
\begin{equation} \label{inflaton_eq}
\ddot\phi+(3H+\Upsilon)\dot\phi +V'(\phi)=0,
\end{equation}
alongside the sourced equation for the radiation energy density,
$\rho_R$,
\begin{equation}
\dot\rho_R+4H\rho_R=\Upsilon\dot\phi^2,
\end{equation}
which can be derived from energy-momentum conservation or from an
explicit computation of particle production rates. The dissipation
coefficient $\Upsilon=\Upsilon(\phi,T)$ can be computed from first
principles using standard thermal field theory techniques, at least
close to thermal equilibrium and in the regime $T\gtrsim H$ where
space-time curvature effects can be neglected (see
Ref.~\cite{Berera:2008ar} for a review).

In the slow-roll regime we then have $\dot\phi \simeq V'/3H(1+Q)$,
where $Q=\Upsilon/3H$ is the dissipative ratio, and $\rho_R/V\simeq
\epsilon_\phi Q /\left[2(1+Q)^2\right]$, provided that the slow-roll
conditions,
%
$\epsilon_\phi,  |\eta_\phi| < 1+Q$
%
are satisfied, where $\epsilon_\phi \equiv M_{\rm
  P}^2\left({V'/V}\right)^2/2$ and  $|\eta_\phi| \equiv M_{\rm
  P}^2{|V''|/V}$ are the standard slow-roll parameters.  Hence, {\it
  if} strong dissipation can be achieved throughout inflation, $Q\gg
1$, the eta-problem is avoided and inflation can occur for inflaton
masses above the Hubble scale~\cite{Berera:1999ws, Berera:2004vm,
  BasteroGil:2009ec}. It is also manifest that radiation is a
sub-dominant component in the slow-roll regime, while it may come to
dominate if $Q$ becomes large at the end of inflation. In warm
inflation there can thus be a smooth transition between inflation and
the radiation era with no need for a separate reheating
period~\cite{Berera:1996fm} where the inflaton decays away,  allowing
e.g.~for cosmic magnetic field generation~\cite{Berera:1998hv},
baryogenesis~\cite{Brandenberger:2003kc,BasteroGil:2011cx} and
inflaton dark matter~\cite{Rosa:2018iff} or dark
energy~\cite{Dimopoulos:2019gpz,Rosa:2019jci,Lima:2019yyv}.

Dissipative effects also lead to thermal inflaton fluctuations,
according to the fluctuation-dissipation
theorem~\cite{Berera:1995wh,Berera:1999ws}, which are sourced by a
thermal noise term with variance $\langle
\xi_\mathbf{k}\xi_\mathbf{k'}\rangle= 2(\Upsilon+H)T/a^3\times
(2\pi)^3\delta^3(\mathbf{k}+\mathbf{k'})$, for $T \gg  H$. Thus,
thermal fluctuations are generically larger than their quantum
counterparts, such that primordial density fluctuations are generated
by classical field perturbations. Moreover, in the strong dissipation
regime,  when $Q  > 1$, their amplitude freezes out before they
become superhorizon~\cite{Berera:1999ws}, nevertheless producing a
nearly scale-invariant spectrum of classical curvature
perturbations~\cite{Hall:2003zp,Graham:2009bf,Ramos:2013nsa}.
{}Finally, such an enhancement of the primordial scalar perturbations
typically leads to a lower inflationary energy scale and, therefore,
to a lower tensor-to-scalar ratio, since gravitational wave production
is unaffected by thermal effects below the Planck scale, as observed
e.g.~in \cite{BasteroGil:2009ec, Cai:2010wt, Bartrum:2013fia}.

Despite all these appealing features, consistently realizing warm
inflation in QFT models has proved to be an enormous
challenge~\cite{Berera:1998gx,Yokoyama:1998ju}. The inflaton typically
gives a large mass to the particles it interacts with, similarly to
the Higgs mechanism, and it is extremely hard to sustain sufficiently
strong dissipative effects due to Boltzmann suppression. Even if this
can be avoided, thermal backreaction generically reintroduces the
eta-problem through thermal inflaton mass corrections $\Delta m_\phi
\sim T \gtrsim H$ up to dimensionless couplings~\footnote{Dissipative
  effects mediated by heavy virtual particles may be sufficient to
  sustain a thermal bath during inflation, but only at the expense of
  very large field multiplicities in supersymmetric
  models~\cite{Berera:2002sp, Moss:2006gt, BasteroGil:2010pb,
    BasteroGil:2012cm}, which may only be found in specific
  constructions in string theory~\cite{BasteroGil:2011mr} or possibly
  other scenarios with extra-dimensions~\cite{Matsuda:2012kc}.}. 

These problems were recently overcome by employing symmetry arguments,
in a model akin to ``Little Higgs" models for electroweak symmetry
breaking and dubbed the ``Warm Little Inflaton" (WLI)
model~\cite{Bastero-Gil:2016qru}. In this model, the inflaton
corresponds to the relative phase between two complex scalar fields,
$\phi_1$ and $\phi_2$, equally charged under a $U(1)$ gauge symmetry,
and which spontaneously break the latter. In the unitary gauge, the
resulting vacuum manifold can be parametrized as
\begin{equation} \label{vacuum}
\langle \phi_1\rangle = {M\over \sqrt{2}}e^{i\phi/M}, \quad \langle
\phi_2\rangle = {M\over \sqrt{2}}e^{-i\phi/M},
\end{equation}
where $M$ is the symmetry breaking scale and $\phi$ is the
gauge-invariant inflaton field\footnote{ 
We emphasize that the relative phase between two equally charged complex scalars does not change under gauge transformations, making our inflaton field distinct from other (pseudo-)Goldstone bosons such as axions used in natural inflaton~\cite{Freese:1990rb} and string monodromy scenarios~\cite{McAllister:2008hb}. Note, in particular, that the $U(1)$ symmetry does not, therefore, constrain the inflaton potential.}. 
In the original model, these complex
scalar fields were coupled to fermion fields $\psi_{1,2}$ through
Yukawa interactions satisfying a discrete interchange symmetry
$\phi_1\leftrightarrow i\phi_2$, $\psi_1\leftrightarrow \psi_2$. This
symmetry then ensures that if e.g.~the $\psi_1$ fermion couples to the
linear combination $\phi_1+\phi_2$, then $\psi_2$ couples to
$\phi_1-\phi_2$. As a result, the fermions acquire masses which are
trigonometric functions of the inflaton field,
\begin{equation} \label{masses}
m_1= gM\cos(\phi/M), \quad m_2=gM\sin(\phi/M),
\end{equation}
where $g$ denotes the Yukawa coupling. These masses are thus bounded
even if $\phi \gg M$, and can be below the temperature during
inflation. Moreover, the leading terms in the finite-temperature
effective potential for $m_{1,2}\lesssim T$,
\begin{equation}
\Delta V_{T} =-{7\pi^2\over 180}T^4+ {1\over 12}
(m_1^2+m_2^2)T^2+\ldots~,
\label{DeltaT}
\end{equation}
are independent of the inflaton field, thus eliminating the
troublesome thermal corrections to the inflaton's mass.

This model
thus yields a consistent realization of warm inflation, and
we have moreover shown that its observational predictions are in
agreement with the Planck data for a quartic inflaton potential,
$V(\phi)= \lambda\phi^4$ \cite{Bastero-Gil:2016qru,Benetti:2016jhf,
  Bastero-Gil:2017wwl, Bastero-Gil:2018uep}. However,
this agreement requires {\it  weak dissipation},  $Q_*\lesssim 1$, at
the time the relevant CMB scales become super-horizon, about 60
e-folds before the end of inflation. Even though $Q$ becomes large
towards the end of the slow-roll regime, the eta-problem remains in
this case.

This is an inherent consequence of the form of the dissipation
coefficient in this scenario, $\Upsilon \propto T$. As originally
shown in \cite{Graham:2009bf} and further analyzed numerically in
\cite{BasteroGil:2011xd}, the temperature dependence of the
dissipation coefficient necessarily leads to a coupling between
inflaton fluctuations and perturbations in the radiation fluid that
modifies the evolution of the former. In particular, for
$d\Upsilon/dT>0$, this results in a substantial enhancement of
inflaton fluctuations if $Q_*\gtrsim 1$. Physically, this is a
consequence of dissipation increasing the temperature more in regions
where it is already higher than average. If $Q$ grows during
inflation, as for the quartic potential, the primordial perturbation
spectrum then becomes blue-tilted, which is ruled out by
Planck~\cite{Akrami:2018odb}.

If, however, $d\Upsilon/dT<0$, inflaton perturbations are damped. More
concretely, the dimensionless curvature power spectrum in the strong
dissipation regime, $Q_*\gg 1$, is generically well approximated by:
\begin{equation} \label{spectrum}
\Delta_\mathcal{R}\simeq {\sqrt{3\pi}\over
  24\pi^2}\epsilon_{\phi_*}^{-1}{V(\phi_*)\over M_{\rm
    P}^4}\left({T_*\over H_*}\right)Q_*^{5/2}\left(Q_*\over
Q_c\right)^{\beta_c},
\end{equation}
where the last factor results  from the interplay between the inflaton
and radiation fluctuations, with $1\lesssim Q_c\lesssim 10$ and
$\beta_c$ depending on $c=d\log \Upsilon /d\log T$ at
horizon-crossing. {}For instance, numerically solving the coupled
system of perturbations as detailed in e.g.~\cite{BasteroGil:2011xd},
we find  $\beta_{1}\simeq 2.5$ for $\Upsilon\propto T$, while for
$\Upsilon\propto T^{-1}$ yields $\beta_{-1}\simeq -1.6$.  Thus, it is
tantamount to find a model with a dissipation coefficient decreasing
with temperature, such as too allow for strong dissipation in an
observationally consistent way.

Here, the complex scalar fields $\phi_{1}$ and $\phi_2$ are coupled to two other complex scalars $\chi_1$ and $\chi_2$ in the thermal bath. The interactions have a renormalizable bi-quadratic form and,
as in the original WLI model, satisfy the discrete interchange
symmetry $\phi_1\leftrightarrow i\phi_2$, $\chi_1\leftrightarrow \chi_2$. Without loss of generality, we may write the relevant
interaction Lagrangian density in the form\footnote{Note that other possible interaction terms involving the gauge-invariant inflaton field may be neglected in a technically natural way.}:
\begin{equation}
  \mathcal{L}_{\phi\chi}=-{1\over2}g^2|\phi_1+\phi_2|^2|\chi_1|^2-
          {1\over2}g^2|\phi_1-\phi_2|^2|\chi_2|^2,
\label{Lphichi}          
\end{equation}
which is a straightforward generalization of the fermionic WLI
model. With the vacuum parametrization in Eq.~(\ref{vacuum}), the
zero-temperature masses of the $\chi_1$ and $\chi_2$ fields are also
given by Eq.~(\ref{masses}), being bounded functions of the inflaton
field $\phi$. Most importantly, their leading contributions to the
inflaton thermal mass also cancel out as in the fermionic case. 
 
The main difference between coupling fermions or scalars to the
inflaton field lies in the form of the dissipation coefficient, due
to their different statistics at non-zero temperature, which will be
fundamental to set the present model apart from previous model
building realizations of warm inflation. {}For on-shell particle
production, which is the dominant process for $T\gg m_{1,2}$, the
dissipation coefficient is given by
\begin{equation}
\Upsilon=\sum_{i=1,2}{g^4\sin^2(2\phi/M)\over 2T}\int {d^3p\over
  (2\pi)^3}{n_B(1+n_B)\over \Gamma_i \omega_{p,i}^2},
\label{Upsilondiss}
\end{equation}
where $n_B(\omega_{p,i})$ is the Bose-Einstein distribution,
$\omega_{p,i}^2=p^2+\tilde{m}_i^2$, $\tilde{m}_i^2=m_i^2+\alpha^2T^2$
is the thermally corrected mass of the $\chi_{1,2}$ fields and
$\Gamma_i$ their thermal decay width. These depend on interactions
within the thermal bath, which we model as Yukawa interactions with
light fermions $\psi_{L,R}$ (with  appropriate charges) and scalar
self-interactions,
\begin{equation}
\mathcal{L}_{\mathrm{\chi\psi}}\!=\!\!\!\!\!\sum_{\substack{i\neq
    j=1,2}}\!\!\!\left(h\chi_i\bar\psi_L\psi_R\!+\!\mathrm{h.c.}
\!-\!{\lambda\over
  2}|\chi_i|^4\!-\!{\lambda'}|\chi_i|^2|\chi_j|^2\!\right)\!.
\label{Lchipsi}
\end{equation}
Scalar self-interactions contribute only at two-loop order to the
thermal decay width, and we focus on parametric regimes where the
decay into light fermions is dominant. Both types of interactions
contribute nevertheless to the thermal mass at the same order,
yielding $\alpha^2\simeq \left[h^2+ 2 \lambda +\lambda' \right]/12$.
The resulting dissipation coefficient is then given by:
\begin{equation} \label{dissip}
\Upsilon\simeq {4 g^4\over h^2}{M^2 T^2\over
  m_\chi^3}\left[1+{1\over\sqrt{2\pi}}\left({m_\chi\over
    T}\right)^{3/2}\right]e^{-m_\chi/T},
\end{equation}
where we have taken the average of the oscillatory terms for field
excursions $\Delta\phi\gg M$, yielding an average mass $m_\chi^2
\simeq g^2M^2/2+\alpha^2T^2$ for both $\chi_i$ scalar fields. 

Although the dissipation coefficient has, in general, a non-trivial
temperature dependence, when $m_\chi$ is dominated by thermal effects
we have $m_\chi\simeq \alpha T$ and $\Upsilon\propto T^{-1}$. This
will then yield the required damping of inflaton fluctuations as
discussed above.

{}Furthermore, the generalization of the 
result for the finite temperature effective potential (\ref{DeltaT}), 
when now accounting for the interactions (\ref{Lphichi})
and (\ref{Lchipsi}) and that can be easily obtained through the
standard means~\cite{Dolan:1973qd,kapusta}, gives
\begin{eqnarray}
&&\!\!\!\!\!\!\!\!\!\!\!\!\!\!\!\!\!\!  \Delta V  \simeq   
- g_*\frac{\pi^2 T^4}{90} \!+\! \sum_{i=1}^2\left\{
\frac{m_{i}^2(\phi) T^2}{12} 
\right.
\nonumber \\
&& \!\!\!\!\!\!\!\!\!\!\!\!\!\!\!\!\!\! - \left.  
\frac{T^4}{6 \pi^2} \left[ \frac{m_i^2(\phi)}{T^2} \right]^{3/2} - {m_i^4(\phi)\over
  32\pi^2}\left[\ln\left({\mu^2\over T^2}\right)-c_b\right] \right\},
  \nonumber \\
\label{DeltaT2}
\end{eqnarray} 
where $\mu$ is a renormalization scale,
$c_b\simeq 3.9076$ and  $g_*$ accounts for the degrees of freedom of the thermal radiation bath,
which, when considering the $\chi_{1,2}$ and $\psi$  fields, 
gives $g_* = 4 + 7/8 \times 4 =7.5$ (for a non-thermalized inflaton),
which is the value we have used in our numerical examples below.   
Adding the contribution of both scalars $\chi_i$, from the
form for their masses $m_{1,2}(\phi)$, which are still given by Eq.~(\ref{masses}),
we immediately see that the
terms that would otherwise contribute as a  quadratic thermal mass
correction to the inflaton, $m_{1}^2(\phi)T^2/12 +
m_{2}^2(\phi)T^2/12 = g^2 M^2 T^2/12$,
become actually
independent of the inflaton field, as we have already anticipated above, 
and it is just a constant term
contributing to the radiation energy density.  
Note also that this cancellation holds beyond the one-loop approximation for the
effective potential, e.g., when resumming self-energy corrections, as the 
so-called ring diagrams~\cite{kapusta}, 
which corresponds
to including the leading thermal corrections to the $\chi_{1,2}$ masses from the Yukawa and $\chi_i$ 
inter- and self-interactions 
given in Eq.~(\ref{Lchipsi}), which gives the additional contribution to the effective
potential,
\begin{equation}
\Delta V_{T,{\rm ring}} = \frac{T^4}{6 \pi} \sum_{i=1,2}  
\left\{ \left[ \frac{m_i^2(\phi)}{T^2} \right]^{3/2}
- \left[ \frac{m_i^2(T)}{T^2} \right]^{3/2} \right\},
\label{Vring}
\end{equation}
where $m_i^2(T) = m_i^2(\phi)+\alpha^2T^2$ is the finite temperature corrected mass for the $\chi_{1,2}$ fields.
Putting together the results from Eqs.~(\ref{DeltaT2}) and (\ref{Vring}), we are then left with only
sub-leading Coleman-Weinberg and radiation terms, with an overall
contribution to the inflaton potential given by 
\begin{eqnarray}
&&\!\!\!\!\!\!\!\!\!\!\!\!\!\!\!\!\!\!  \Delta V_T \simeq -g_* \frac{\pi^2 T^4}{90} + g^2 M^2 \frac{T^2}{12} 
\nonumber \\
&& \!\!\!\!\!\!\!\!\!\!\!\!\!\!\!\!\!\! -
\frac{(gM)^4 \left[\cos^4\left(\frac{\phi}{M}\right) + \sin^4\left(\frac{\phi}{M}\right)
    \right]}{32\pi^2}\left[\ln\left({\mu^2\over
    T^2}\right)-c_b\right]
\nonumber \\
&&\!\!\!\!\!\!\!\!\!\!\!\!\!\!\!\!\!\!  - \frac{T}{6 \pi} \left[
g^2 M^2 \cos^2\left(\frac{\phi}{M}\right) + \alpha^2 T^2 \right]^{3/2}
\nonumber \\
&&\!\!\!\!\!\!\!\!\!\!\!\!\!\!\!\!\!\!  - \frac{T}{6 \pi} \left[
g^2 M^2 \sin^2\left(\frac{\phi}{M}\right) + \alpha^2 T^2 \right]^{3/2}.
\label{DeltaV}
\end{eqnarray} 
We also note that the renormalization scale dependence of the
effective potential is, of course, an artifact of working at finite order in perturbation theory and, 
as standard in the literature, we can choose $\mu=T_*$ (besides of also absorbing the constant $c_b$
into the definition of $\mu$ inside the log term) in order to minimize this
dependence at horizon-crossing when the relevant observables are measured.
Since $\Delta V_T \sim \rho_{R} < \rho_\phi$ and with an explicit potential $V$ for the inflaton,
one always has $|\Delta V_T|/V  \ll 1$ during inflation as long as the slow-roll conditions are satisfied.
The $\phi$ dependence on Eq.~(\ref{DeltaV}) are only through highly oscillatory (and bounded terms),
since $\phi \gg M$, for the relevant parameters of the model (see below).
We have nevertheless explicitly checked and confirmed numerically
that these oscillatory terms do not significantly affect the dynamics of inflation.
This result is also very much analogous to the one found in the earlier model introduced
in Ref.~\cite{Bastero-Gil:2016qru}.

Let us consider the simplest scenario with a quadratic inflaton
potential\footnote{Strictly speaking the interchange symmetry implies
  $V(\phi)\propto (\phi/M-\pi/4)^2$, but for $\phi\gg M$ we may take
  $V(\phi)\propto \phi^2$ for simplicity.},
$V(\phi)=m_\phi^2\phi^2/2$, since as we discussed above there are no
symmetries protecting the inflaton from acquiring at least a
Hubble-scale mass. The slow-roll equations can be integrated
analytically for this potential when $\Upsilon\propto T^{-1}$, and
from the form of the primordial perturbation spectrum at strong
dissipation in Eq.~(\ref{spectrum}) we obtain for the scalar spectral
index $n_s-1\simeq 2\beta_{-1}/N_e$. While this actually gives a too
red-tilted spectrum $n_s\simeq 0.95$ for $N_e=60$ e-folds of
inflation, the zero-temperature mass of the $\chi_i$ fields
generically leads to non-negligible deviations from $\Upsilon\propto
T^{-1}$ that make the spectrum more blue-tilted. 

\begin{figure}[!t]
\includegraphics[width=7.5cm]{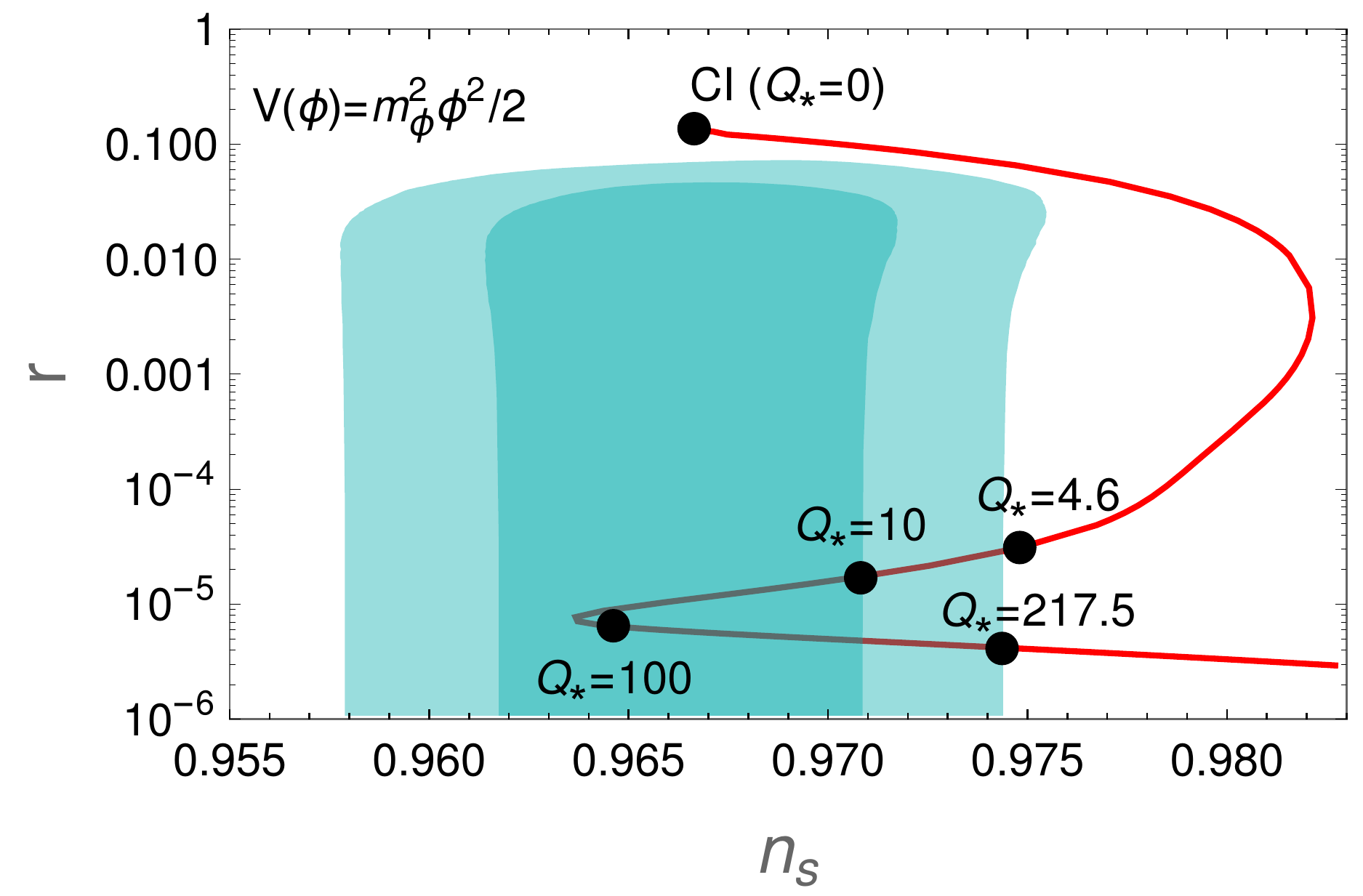}
\caption{Observational predictions with a quadratic inflaton
  potential,  varying the dissipative ratio $Q_*$ at horizon-crossing
  60 e-folds before the end of inflation, as indicated by the black
  circles. The model parameters used are
      explained in the text.  The cold inflation (CI) prediction is
  also indicated. The shaded regions are for the $68\%$ and $95\%$
  C.L. results from the Planck 2018 legacy data
  (TT+TE+EE+lowE+lensing+BK15+BAO)~\cite{Akrami:2018odb}.}
\label{fig1}
\end{figure}

We have then solved numerically the background and perturbation
equations for both the inflaton field and radiation fluid with the
full form of the dissipation coefficient in Eq.~(\ref{dissip}),
following the procedure described in detail in
e.g.~\cite{BasteroGil:2011xd}. In {}Fig.~\ref{fig1}, we show the
observational predictions in the interesting parametric regime where
the mass of the $\chi_i$ fields is dominated by thermal
effects. To obtain  the results shown in
    {}Fig.~\ref{fig1},  the product $gM$ appearing in the dissipation
    coefficient~(\ref{dissip}) was kept fixed in the value
    $gM=2.6\times10^{-5}M_{\rm P}$ and we also considered $\alpha^2=1/8$ for
    definiteness. Variations in these parameters tend only to shift the
    values of $Q_*$ with respect to the ones seen in
    {}Fig.~\ref{fig1}.  We have controlled the magnitude of the
    dissipation ratio $Q_*$ through the ratio of the coupling
    constants $g^2/h^2$ in the prefactor of (\ref{dissip}).  We have
    also fixed the amplitude of the curvature power spectrum as
    $\Delta_\mathcal{R} = 2.2 \times 10^{-10}$, consistent with the
    Planck legacy data~\cite{Akrami:2018odb}.  This allows one to
    obtain the mass parameter $m_\phi$ of the inflaton potential in a
    consistent way for each value of $Q_*$, using, e.g.,
    Eq.~(\ref{spectrum}).

As one can see in {}Fig.~\ref{fig1}, for dissipation ratio values at
horizon-crossing $Q_*\lesssim 4.6$, the spectral tilt is too large,
exceeding the corresponding cold inflation prediction. This is due to
the thermal nature of inflaton fluctuations and the increase of the
dissipative ratio $Q$ during inflation. As we increase $Q_*$, however,
the spectrum becomes more red-tilted as a consequence of the damping
induced by the stronger coupling with radiation fluctuations near
horizon-crossing. {}For $Q_*\sim 10-180$, we in fact find a spectral
tilt $n_s$ in very good agreement with the Planck $68\%$ C.L. results
and a low tensor-to-scalar ratio $r\lesssim 10^{-5}$, as typical of
warm inflation models. In particular, for the case of $Q_*=100$ and
parameters shown in Fig.~\ref{fig1}, we have that $n_s\simeq 0.965$
and $r \simeq 6.4 \times 10^{-6}$. 
  
  In this strong dissipative regime, primordial non-Gaussianity should
  generically be at the level $|f_{\rm NL}^{\text{warm}}|\lesssim 10$
  \cite{Moss:2007cv}, and the dedicated searches by the Planck
  collaboration for the warm shape of the bispectrum
  \cite{Ade:2015ava, Akrami:2019izv} allow for $Q_*< (3.2-4)\times
  10^3 $ (95\% C.L.). Our scenario thus lies comfortably within these
  limits. We have, moreover, made a preliminary analysis of
  non-Gaussianity for the particular form of the dissipation
  coefficient in the present scenario, using the numerical codes
  developed in \cite{Bastero-Gil:2014raa}, obtaining $f_{\rm
    NL}^{\text{warm}}\simeq 3$ for $Q_*=100$.

Scenarios with  $Q_* \gg 100$ within this model typically require
larger values of the coupling $g$ and hence larger zero-temperature
$\chi_i$ masses. This results in larger deviations from
$\Upsilon\propto T^{-1}$ and thus to a less efficient damping of
inflaton fluctuations and a more blue-tilted spectrum. 

\begin{figure}[!htb]
\includegraphics[width=7.5cm]{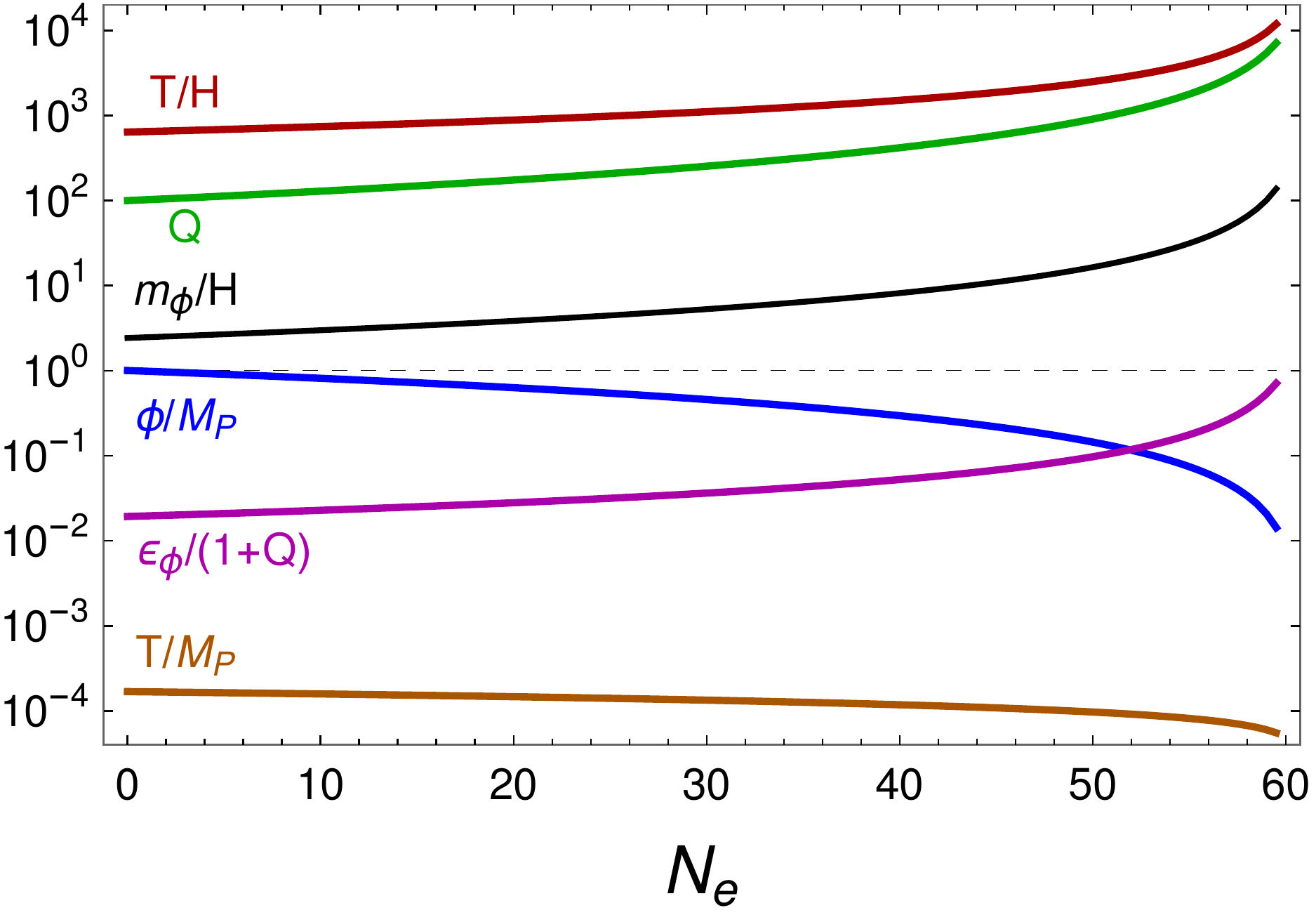}
\caption{Dynamical evolution of different quantities and for the
  particular case of $Q_*=100$.The full set of
      model parameters is given in the text.}
\label{fig2}
\end{figure}

The agreement with observational data for $Q_*\sim 100$ is, however,
our most significant result, since in this regime the slow-roll
parameters $\epsilon_\phi= \eta_\phi\gtrsim 1$ throughout
inflation. The slow-roll trajectory is thus fully sustained by the
dissipative friction, yielding a consistent effective field theory for
inflation with $m_\phi \gtrsim H$, $\Delta\phi \lesssim M_{\rm
  P}$ \footnote{The swampland distance conjecture states that a tower
  of massive states become exponentially light as
  $e^{-\alpha\phi/M_P}$, where $\alpha\sim \mathcal{O}(1)$, for $\phi>
  (1-2)M_P$ (see e.g.~\cite{Palti:2019pca}). In our model, we obtain
  $\phi_*\lesssim M_P$ at horizon-crossing for $Q_*\sim 100-200$ in
  the observationally viable window, which does not constitute a
  significant mass suppression, such that these states may
  consistently be integrated out in the effective field theory.} and
$M_{\rm P}V'/V\gtrsim 1$. We illustrate the dynamical evolution of the
inflaton-radiation system in a representative case with $Q_*=100$ in
{}Fig.~\ref{fig2}, where it is manifest that inflation can
consistently occur with a super-Hubble inflaton mass and a
sub-Planckian field excursion.  As explained
    above, for the choice of parameters for {}Fig.~\ref{fig1},  given
    the value of $Q_*=100$, we have the full set of explicit model
    parameters that reproduces {}Fig.~\ref{fig2} as given by $g\simeq
    0.47$, $M \simeq5.6\times10^{-5}M_{\rm P}$, $\alpha^2=1/8$,
    $m_\phi\simeq 6.4\times10^{-7}M_{\rm P}$ and the couplings in
    (\ref{Lchipsi}) given by  $h\simeq 0.82$, $\lambda \simeq 0.28$
    and we have also chosen $\lambda'=\lambda$.

  In the example shown in {}Fig.~\ref{fig2}, it is also manifest that
  strong dissipation can be sustained for a whole of 60 e-folds of
  inflation, maintaining a slowly decreasing
  temperature\footnote{Although not explicitly shown, we have checked
    that the average thermal decay width of the $\chi_i$ fields
    exceeds the Hubble rate, keeping the $\chi_i$ fields and their
    decay products close to equilibrium, and that sub-leading
    corrections to the finite-temperature effective potential, which
    give oscillatory contributions to the slow-roll parameters, have
    no significant effect at both the background and perturbation
    levels.}  $T\gg H$. The temperature always satisfies $T \gtrsim g
  M$, ensuring that the $\chi_i$ fields are relativistic, yet keeping
  the $U(1)$ gauge symmetry broken throughout inflation\footnote{The
    exact critical temperature of the phase transition depends on the
    self-couplings of the $\phi_1$ and $\phi_2$ scalar fields which we
    have not specified, but we can assume $T_c\sim M$ within order
    unity factors.}. The radiation abundance is  $\rho_R/V\simeq
  0.5\epsilon_\phi/Q$ in the slow-roll regime for $Q\gg 1$,  leading
  to a smooth transition to the radiation-dominated era. We note that
  this model also allows for a much shorter inflation period,
  successfully setting all the large-scale observables, along the
  lines of the recently proposed multi-stage warm inflation solution
  to the TCC~\cite{Berera:2019zdd,Berera:2020dvn}.

Hence, this letter shows, for the first time, that slow-roll inflation
does not require an unnaturally light inflaton scalar field, within a
simple renormalizable QFT with a quadratic scalar potential, that is
robust against corrections from unknown new physics, particularly
Planck-suppressed non-renormalizable operators.  Because this model is 
renormalizable in the conventional sense, it is fully decoupled from the
unknown high energy physics. Moreover, the spectrum
of primordial density fluctuations is fully described by classical
thermal fluctuations of the inflaton field that freeze out before
becoming super-horizon, being free of ambiguities in the choice of the
quantum vacuum state, which could moreover be sensitive to unknown
transplanckian effects.

 This model is very simple, involving only a few scalar and fermion
 fields, interacting via renormalizable couplings in the perturbative
 regime. Moreover, the fundamental scale $M\ll M_P$, in line with the
 weak gravity conjecture and with minimal reliance on unknown
 gravitational physics at the Planck scale. The model involves only a
 $U(1)$ symmetry alongside a discrete symmetry, which are ubiquitous
 in Nature. Like the Standard Model, this is truly a low energy model
 that can make reliable predictions without any knowledge of the
 high-energy completion. In particular, this model does not need
 supersymmetry to control quantum corrections, although a
 supersymmetric extension could easily be implemented should this be a
 feature of quantum gravity.

It had been pointed out early in the development of warm
inflation~\cite{Berera:1999ws,Berera:2004vm} that in the strong
dissipation regime it is possible for $m_\phi > H$ to solve the
eta-problem, with sub-Planckian field excursions, thus solving the
swampland criteria well before they were stated.  It had also been
understood that increasing dissipation lowers the energy scale of
inflation and hence the tensor-to-scalar ratio
\cite{Berera:1999ws,BasteroGil:2009ec} in line with subsequent CMB
observations. However, the challenge has been to find a theoretically
and observationally consistent model displaying all these appealing
features.  This work has achieved this by obtaining a QFT model of
inflation that is reliable in this respect, involving only a few
fields, being free of the fine-tuning and ambiguities that generically
plague the more conventional cold scenario.

\section*{Acknowledgments}

A.~B. is partially supported by STFC. \break M.B.-G. is partially supported by
MINECO grant FIS2016-78198-P and Junta de Andaluc\'ia grant SOMM/17/6104/UGR and FQM101.
R.O.R. is partially supported by research grants from Conselho
Nacional de Desenvolvimento Cient\'{\i}fico e Tecnol\'ogico (CNPq),
Grant No. 302545/2017-4, and Funda\c{c}\~ao Carlos Chagas Filho de
Amparo \`a Pesquisa do Estado do Rio de Janeiro (FAPERJ), Grant
No. E-26/202.892/2017. 
J.\,G.\,R. is supported by the FCT Grant No. IF/01597/2015, 
the CFisUC strategic project No. UID/FIS/04564/2019 and partially by the FCT project
PTDC/FIS-OUT/28407/2017 and ENGAGE SKA (POCI-01-0145-FEDER-022217).

\bibliographystyle{elsarticle-num}

\end{document}